\documentstyle[12pt,epsf,epsfig]{article}
\begin{document}
\vspace{5mm} \par \noindent
  \par
\vspace{5mm} \par \noindent
\renewcommand{\thebibliography}[1]{ {\vspace{5mm}\par \noindent{\bf
References}\par \vspace{2mm}}
\list
 {\arabic{enumi}.}{\settowidth\labelwidth{[#1]}\leftmargin\labelwidth
 \advance\leftmargin\labelsep\addtolength{\topsep}{-4em}
 \usecounter{enumi}}
 \def\newblock{\hskip .11em plus .33em minus .07em}
\sloppy\clubpenalty4000\widowpenalty4000
 \sfcode`\.=1000\relax \setlength{\itemsep}{-0.4em} }
\newcommand\rf[1]{(\ref{#1})}
\def\nn{\nonumber}
\newcommand{\ft}[2]{{\textstyle\frac{#1}{#2}}}

\thispagestyle{empty}

\begin{center}

\vspace{3cm}

{\large\bf Entropy Bounds and String Cosmology}\\

\vspace{1.4cm}

{\sc G. Veneziano}\\

\vspace{1.3cm}

{\em Theoretical Physics Division, CERN} \\
{\em CH-1211 Geneva 23, Switzerland} \\

\vspace{1.2cm}

\centerline{\bf Abstract}
\vspace{- 4 mm}  \end{center}
\begin{quote}\small
After discussing some old (and not-so-old) entropy bounds both for isolated systems and
in cosmology, I will argue in favour of a ``Hubble entropy bound"  holding
in the latter context.
I will then apply this bound to recent developments in string cosmology, show that it is
naturally saturated throughout pre-big bang inflation, and claim that
 its fulfilment at later times
has interesting implications for the exit problem of string cosmology.
 
\end{quote}
\vfill

\def\obdot{\hskip-8pt \vbox to 11pt{\hbox{..}\vfill}}
\def\obbdot{\hskip-8pt \vbox to 14pt{\hbox{..}\vfill}}
\def\odot{\hskip-6pt \vbox to 6pt{\hbox{..}\vfill}}

\def\laq{\raise 0.4ex\hbox{$<$}\kern -0.8em\lower 0.62
ex\hbox{$\sim$}}
\def\gaq{\raise 0.4ex\hbox{$>$}\kern -0.7em\lower 0.62
ex\hbox{$\sim$}}
\def\beq{\begin{equation}}
\def\eeq{\end{equation}}
\def\bea{\begin{eqnarray}}
\def\eea{\end{eqnarray}}
\def\bean{\begin{eqnarray*}}
\def\eean{\end{eqnarray*}}

\def \bk {{\bf k}}
\def \pa {\partial}
\def \ra {\rightarrow}
\def \fb {\overline \phi}
\def \fbp {\dot{\fb}}
\def \bp {\dot{\beta}}
\def \rb {\overline \rho}
\def \pb {\overline p}
\def \pr {\prime}
\def \se {\prime \prime}
\def \H {{a^\prime \over a}}
\def \fp {{\phi^\prime}}
\def \ti {\tilde}
\def \la {\lambda}
\def \ls {\lambda_s}
\def \La {\Lambda}
\def \Da {\Delta}
\def \b {\beta}
\def \a {\alpha}
\def \ap {\alpha^{\prime}}
\def \ka {\kappa}
\def \Ga {\Gamma}
\def \ga {\gamma}
\def \sg {\sigma}
\def \da {\delta}
\def \ep {\epsilon}
\def \r {\rho}
\def \om {\omega}
\def \Om {\Omega}
\def \noi {\noindent}
\def \pfb {\Pi_{\fb}}
\def \pM {\Pi_{M}}
\def \pbe {\Pi_{\b}}
\def \pa {\partial}
\def \dd {\partial}
\def \ra {\rightarrow}
\def \fb {\overline \phi}
\def \fbp {\dot{\fb}}
\def \bp {\dot{\beta}}
\def \rb {\overline \rho}
\def \pb {\overline p}
\def \pr {\prime}
\def \se {\prime \prime}
\def \H {{a^\prime \over a}}
\def \fp {{\phi^\prime}}
\def \ti {\tilde}
\def \al {\alpha}
\def \la {\lambda}
\def \ls {\lambda_s}
\def \La {\Lambda}
\def \Da {\Delta}
\def \De {\Delta}
\def \de {\delta}
\def \b {\beta}
\def \a {\alpha}
\def \ap {\alpha^{\prime}}
\def \ka {\kappa}
\def \Ga {\Gamma}
\def \ga {\gamma}
\def \sg {\sigma}
\def \Sg {\Sigma}
\def \si {\sigma}
\def \da {\delta}
\def \ep {\epsilon}
\def \r {\rho}
\def \om {\omega}
\def \Om {\Omega}
\def \noi {\noindent}
\def \pfb {\Pi_{\fb}}
\def \pM {\Pi_{M}}
\def \pbe {\Pi_{\b}}
\def \lap {\triangle}














Why is the second law of thermodynamics valid even when the microscopic evolution equations
are invariant under time-reversal?
The standard answer to this old question (see e.g. \cite{Penrose}) is simple:
 it is because the Universe started in a low-entropy state and has not yet reached   its
 maximal attainable entropy. But then, which is this maximal possible value of entropy
and why has it not already been reached after so many billion years of cosmic evolution?
In this talk I will argue that, perhaps, there is a simple answer to these last two
questions, at least in the context of string cosmology. But let us proceed step by step.

In 1981 J. Bekenstein \cite{BB} proposed what he called a ``universal" entropy bound
for isolated objects. We will refer to it as the Bekenstein entropy bound (BEB) \cite{BB},
 which states that, for any isolated physical system of energy $E$
and size $R$, usual thermodynamic entropy is bound by \footnote{Throughout
this paper we stress functional dependences
while ignoring numerical factors and set $c=1$.}:
\begin{equation}
 S \le S_{BB} = ER/\hbar.
\label{BE}
\end{equation}
 I will skip the arguments that led
Bekenstein to formulate his bound and just stress that, in 18 years, no counterexample to it has been
found. 

The so-called 
holographic principle of 't Hooft, Susskind and others \cite{holography}, suggests
an apparently unrelated holographic bound 
on entropy (HOEB)  according to which entropy cannot
exceed one unit per Planckian area of its boundary's surface. In formulae:
\begin{equation}
 S \le S_{HOB} = A ~ l_P^{-2}.
\label{HO}
\end{equation}

I will now argue that the BEB actually implies the HOEB. Indeed:
\begin{equation}
 S_{BB} = GER/G\hbar = R_s~ R ~l_P^{-2} \le S_{HOB} = R_{eff}^2 ~ l_P^{-2} \;\;, \; R_s \equiv GE \; ,
\label{comparison}
\end{equation}
where $R_{eff}$ appearing in the holography bound is $R$ if $R>R_s$ (a non-collapsed object),
but has to be identified with
 $R_s$ if the object is  inside its own Schwarzschild radius (is itself a black hole).
In the latter  case the two bounds coincide and are saturated.

Incidentally, the BEB has an amusing application \cite{GVDivonne}
 to (weakly coupled) string theory. 
Since string entropy is  $O(\alpha' E / l_s)$ (one unit
 per string length $l_s = \sqrt{\alpha' \hbar}$),
 it satisfies the BEB
iff $R > l_s$. Thus, in string theory, one cannot have black holes with Schwarzschild radius smaller
than $l_s$ (with a Hawking temperature larger than the string's Hagedorn temperature) \cite{GVFC}. 

The situation for isolated systems in flat space-time looks  uncontroversial.
How can we try to extend these considerations to a cosmological set up?
Let us first pretend that we can use the naive BEB or holography bounds to an arbitrary
sphere of radius $R$, cut out of a homogeneous cosmological space. Entropy in cosmology is
extensive, i.e. it grows like $R^3$. But the boundary's
  area grows  like $R^2$: therefore, at sufficiently large $R$,
  the (naive) holography bound must be violated! On the other hand, $S_{BB} \sim E R \sim
R^4$ appears to be safer at large $R$. How can this be, since we just argued that the BEB
implies the HOEB? The explanation is simple: when $R$ becomes very large, the corresponding
 $R_s$  exceeds $R$; nevertheless,
 we kept using $R$ in the HOEB since we  no longer had a black-hole interpretation
for the sphere. Obviously, we have to rethink everything within a cosmological setting!

In order to show how inadequate the naive bounds are in cosmology, let us apply them
at $t \sim t_P \sim 10^{-43}~{\rm s}$, within standard  
 FRW cosmology, to the region of space that
 has become our visible Universe today. The size of that region at $t \sim t_P$ 
was about $10^{30} l_P$
and the entropy density was of Planckian order. Thus:
\begin{eqnarray}
&&S \sim (10^{30})^3 = 10^{90} \;\; ,\\ \nonumber
  &&S_{BB} \sim \rho R^4/\hbar \sim R^4/l_P^4 \sim 10^{120} \;\;,
 S_{HOB} \sim R^2 l_P^{-2} \sim 10^{60}\; .
\end{eqnarray} 
Clearly the actual entropy lies at the geometric mean
 between the two naive bounds, making one false
 and the other quite useless!

It was indeed realized by their respective proponents that both the BEB and the HOEB
need revision in a cosmological context. In 1989 Bekenstein proposed \cite{Bekenstein2}
that the BEB applies to a region as large as the particle horizon $d_p$:
\begin{equation}
d_p(t) = a(t) \int_{t_{beg}}^t dt'/a(t') \; .
\label{dp}
\end{equation}
The same conclusion (with an important caveat, see below) was
 reached by Fischler and Susskind \cite{FS} (FS) in their cosmological
generalization of the HOEB. 

There is one very welcome property of both the Cosmological BEB and the FS bound: 
they appear to be saturated around the 
Planck time (when they can be shown to be equivalent) and could thus justify the initially ``low"
entropy value.
Actually,  one finds \cite{Bekenstein2} that the bound is saturated at $t \sim N_{eff}^{1/2}~t_p$
and is violated at earlier times if one trusts General Relativity so far inside the strong-curvature
region. This result
was used by Bekenstein \cite{Bekenstein2}
to argue that the Big Bang singularity must be spurious.

It is interesting to compare
 the two bounds again, now in their cosmological variants. They are related as follows:
\begin{equation}
 S_{CBB} \sim M(r< d_p) d_p/\hbar = \rho d_p^4/\hbar = (H d_p)^2  d_p^2/l_p^2 =
 (H d_p)^2 S_{CHOB} \; ,
\label{CosmBs}
\end{equation}
where, with increasingly baroque notation, we have added a $C$ to distinguish the cosmological 
versions of the two bounds and we have used Friedmann's equation $G \rho = H^2$ to relate energy
density to the Hubble parameter $H = \dot{a}/a$.

We note that the two bounds differ by a factor $(H d_p)^2$. While such a factor is $O(1)$
in FRW-type cosmologies, it can be huge after a long period of inflation, i.e.
$O\left((a_{end}/a_{beg})^2\right)$, the square of the total amount of red-shift 
suffered during inflation, 
which has to be at least as large as $10^{60}$.
For this reason the CHOEB (FSB hereafter) appears to be stronger than the CBEB, just the opposite
of what we argued to be the case for isolated systems.

 The tight nature of
the FSB led some authors \cite {Rama} to derive constraints  from it on inflationary parameters.
This, however, came from a misinterpretation of the FSB \footnote{This point was clarified
after my talk through several discussions with Fischler and Susskind, see also Ref. \cite{KL}.}.
The logical implication of the FSB is that it does not apply to entropy 
produced by non-adiabatic processes occurring in the bulk. In any inflationary scenario,
most of the present entropy is the result of processes of this type (reheating due to
dissipation of the inflaton's potential energy  at the end of inflation \cite{inflation}) 
and should therefore be excluded. As a result, 
the FSB puts no constraints on inflation, but also becomes phenomenologically 
uninteresting in recent epochs, since it ignores most of the present entropy. On the contrary, the FSB
appears to exclude closed, recollapsing universes \cite{FS}, or those driven by a small negative
cosmological constant \cite{KL}.

Two groups \cite{Rey1}, \cite{Maharana} tried to apply the FSB to pre-big bang (PBB) cosmology.
A problem arises, however, since the particle horizon $d_p$ is infinite in PBB (the integral
in Eq. (\ref{dp}) diverges at its lower limit, $- \infty$). One of the groups \cite{Rey1} insisted on
using $d_p$ nonetheless, and  concluded that the PBB initial state has to be
empty. The second group \cite{Maharana} replaced the particle horizon with the
event horizon (which is finite in  PBB and infinite in FRW) and found very mild constraints.
Very recently, Bousso \cite{Bousso}  proposed to change the FS prescription by replacing
 $d_p$
with yet another scale, and thus managed to avoid the above-mentioned
 problems \cite{FS} with a recollapsing universe.
In the rest of this talk I will argue in favour of a different cosmological entropy bound,
which is unambiguous and appears to give sensible results. I will then apply it to the 
PBB scenario.

Consider a  sufficiently homogeneous Universe with its (local, time-dependent) Hubble 
expansion  rate defined, in the synchronous gauge, by:
\begin{equation}
 6 H = - 2 K \equiv  \partial_t~(\log g) ~~ , ~~ g \equiv {\rm det}~(g_{ij}) ~~,
\label{defH}
\end{equation}
where, as usual, $K$ denotes the trace of the second fundamental form on constant $t$
 hypersurfaces. We assume $H$ to vary little (percentage-wise) 
over distances $O(H^{-1})$. In this case $H^{-1}$, 
the so-called Hubble radius,
 corresponds to the scale of causal connection, i.e. to the scale within which
microphysics can act. 

As long as we consider, on top of this homogeneous background, isolated lumps
of size much smaller that $H^{-1}$, the expansion of the Universe is irrelevant, 
and we should fall back
on the non-cosmological, asymptotically flat case. In particular, we can imagine to
put, in a single Hubble patch, several black holes and compute their entropy. We can
make them coalesce and watch the consequent entropy increase (mass adds up,
but entropy is proportional its square). However, this way of increasing entropy
has some limit. It is hard to imagine that a black hole 
larger than $H^{-1}$ can  form, since different parts of its horizon would
be unable to  hold together. Actually, strong arguments in this direction were
given long ago in the literature \cite{Carr} (see also \cite{KL}). 
Thus, the largest entropy we may conceive 
for a  region of space  larger than $H^{-1}$ is the one
 corresponding to  one black hole per Hubble volume $H^{-3}$. Using
the Bekenstein--Hawking formula for the entropy of a black hole of size $H^{-1}$ leads to the
proposal \cite{GVHB}, \cite{others} of a  ``Hubble entropy bound" (HEB):
\begin{equation}
S(V) < S_{HB} \equiv  n_H S_H = V H^3 l_P^{-2}  H^{-2} = V H l_P^{-2} ~~,
\label{HB}
\end{equation}
where  $n_H$ is the number of Hubble-size
regions within the volume $V$, each one carrying maximal
 entropy $S_H = l_P^{-2}  H^{-2}$. A possible relation
between the HEB and a generalized second law of thermodynamics has also been discussed \cite{Ramy}.

 Note that the bound (\ref{HB})
is partly holographic (the $S_H$ part goes like an area) and partly extensive (the $n_H$ part
goes like the volume). If we apply the HEB to a region of size $d_p$ we find, amusingly:
\begin{equation}
S_{HB} = d_p^3 H l_P^{-2} = S_{CBB}^{1/2}~S_{FS}^{1/2}~~.
\label{HBvsBFS}
\end{equation}
It is easy to show \cite{GVHB} that the above relation is sufficient to avoid any problem with
entropy produced at reheating after inflation. Also, the HEB coincides with the CBEB and FSB
at Planckian times in FRW cosmology and it is thus  as saturated as they are.
In the rest of this talk I will concentrate on applying the HEB to the PBB scenario, showing that, 
in that context, the above saturation is not accidental.

In order to discuss various forms of entropy in the PBB scenario, let us recall some results
that have emerged from recent studies \cite{GV,BDV}
 of the question of initial conditions in string cosmology (see \cite{Erice} for a recent review).
It has been argued that  very natural initial conditions, corresponding
to generic  gravitational and dilatonic waves superimposed on the trivial, perturbative
vacuum of critical superstring theory (flat space-time and constant dilaton),
lead to a form of stochastic PBB. In the Einstein-frame metric, this can be seen as a
 sort of chaotic gravitational collapse which is bound to occur, owing to gravitational
instability through the Hawking--Penrose theorems \cite{HP}, provided a 
(scale and dilaton shift invariant) collapse criterion
is met. Black holes of different
 sizes will thus form but, for an observer measuring distances
inside each black hole with a stringy meter, this is experienced as
 a PBB inflationary cosmology in which 
the (hopefully fake) $t=0$  big bang singularity is identified 
with the (hopefully equally fake)
black hole singularity at $r=0$  \cite{BDV}. Since the duration (and efficiency) of the inflationary
phase is controlled by the size of the black hole,
we are  led \cite{GV,BDV} to identify our observable Universe with what became of a portion of space
that was originally inside a sufficiently large  black hole.

It is helpful to follow the evolution of various contributions to (and bounds on)
entropy with the help of Fig. 1.
At time $t=t_i$, corresponding to the first appearance of a horizon, we can use the
Bekenstein--Hawking formula to argue that
\begin{equation}
 S_{coll} \sim (R_{in}/l_{P,in})^2 \sim (H_{in} l_{P,in})^{-2}  = S_{HB}  \; ,
\label{initialS}
\end{equation}
where we have used the fact \cite{BDV} that the initial size of the black-hole horizon determines
also the initial value of the Hubble parameter.
Thus, at the onset of collapse/inflation,  entropy, without any fine-tuning,
 is as large as allowed by the HEB. As a confirmation of this, note that
  $S_{coll}$ is also on the order of the number of quanta needed for
 collapse to occur \cite{BDV}.
  We have implicitly assumed
the initial  state of the system to
be at small string coupling: consequently, quantum 
fluctuations  are very small,
and contribute,  initially, a negligible amount $S_{qf}$ to the total entropy.

After a short transient phase, dilaton-driven inflation (DDI) should follow
\cite{GV,BDV} and last until $t_s$,
the time at which a string-scale curvature is reached. We expect this classical
 process not
 to generate further entropy (unless more energy flows into the black hole,
 but this would only increase its total comoving volume),
 but what happens
to the HEB? Well, it stays constant, thus keeping the bound saturated,
 as the result of a well known ``conservation law"
of string cosmology \cite{PBB}, which reads $(l^2_P = e^\phi l^2_s)$
 \begin{equation}
\partial_t \left(e^{-\phi} \sqrt{g} H \right) = 
\partial_t \left((\sqrt{g} H^3)~~( e^{-\phi} H^{-2}) \right) = \partial_t \left( n_H S_H \right) = 0~.
\label{conservation}
\end{equation}
 Comparing this  with (\ref{HB}), we 
recognize that (\ref{conservation}) simply expresses the time independence
of the HEB during the DDI phase. At the
beginning of the DDI phase $n_H =1$, and the
whole entropy is in a single Hubble volume; however, as DDI proceeds,
 the same total amount of entropy
becomes equally shared between very many Hubble volumes until, eventually, each one of them
contributes a small number.
Also, if we assume  that the string coupling is still small at the end of DDI, 
we can easily argue that
the entropy in quantum fluctuations remains at a negligible level during that phase.

Is this going to continue indefinitely? Hopefully not: we want to exit from
the DDI phase and enter, eventually, some kind of FRW cosmology! This is the well-known
exit problem of string cosmology \cite{exit}. Two diagrams can be helpful
when discussing this problem. In Fig. 2 we plot, on a linear scale, the Hubble
parameter against a (duality-invariant) combination of the rate of growth of the dilaton and
$H$. DDI lies in the first quadrant of this plane, FRW in the second. If exit occurs, the two
branches should smoothly connect (dotted line). In Fig. 3, we show instead, on a log-log
plot, $H$ as a function of the string coupling. DDI solutions now correspond to the parallel
straight lines going upwards  to the right. Different straight lines correspond to
different initial conditions (different classical moduli). The horizontal
boundary corresponds to the reach of string-scale curvatures, where $\alpha'$ corrections 
should become essential in order to prevent the singularity. 

Let us assume
for the moment initial conditions such that we hit this
boundary while the coupling is still small and ask whether the HEB may
come to our help.
In fact, since the HEB is  saturated all the time during DDI, it cannot decrease
after this phase ends. This condition reads:
\begin{equation}
\partial_t(e^{-\phi} \sqrt{g} H) \ge 0  \Rightarrow (\dot{\phi} - 3H) \le \dot{H}/H \; .
\label{HEBgrowth}
\end{equation}
This constraint is very welcome. As $\alpha'$ corrections intervene to stop the growth
of $H$, the HEB forces $\dot{\phi} - 3H$ to decrease and even to change sign if
$H$ stops growing. But this is
 just what is needed  to make the DDI branch flow into the FRW branch
in Fig. 2!

Consider now the second possibility \cite{MR}, the case in which strong coupling
is reached first, i.e. while the curvature is still small in string units.
In this case we can neglect $\alpha'$ corrections but not loop corrections,
particle production, and back-reaction effects.
When will exit occur? It has been assumed \cite{BBB} that it does when the energy
 in the quantum fluctuations (which can be easily estimated \cite{PBB})
becomes critical, i.e. when 

\begin{equation}
\rho_{qf}  \sim N_{eff} ~ H^4_{max} = \rho_{cr} = e^{-\phi_{exit}} M_s^2 ~ H^2_{max} ~,
\label{BBB}
\end{equation}
where $N_{eff}$ is the effective number of
particle species produced. This gives the exit condition 
$ N_{eff} ~ e^{\phi_{exit}}~l_s^2~H^2_{max} = 1$, 
i.e. the rightmost boundary in Fig. 3. 
Let us show  that this is also the boundary where the HEB is saturated. Using known results 
on entropy production due to the cosmological squeezing of vacuum fluctuations \cite{entropy},
and the previous constraint,
we find:
\begin{equation}
S_{qf}^{({\rm ex})}  \sim N_{eff} ~ H^3_{max} V \sim e^{-\phi_{exit}} l_s^{-2} ~ H_{max} V \sim 
 l_P^{-2} ~ H_{max} V \sim S_{HB}^{({\rm ex})} ~. 
\label{qfentropy}
\end{equation}
Note that the existence of this boundary can also be argued for \cite{MR} on the basis
of $M$-theory:  Kaluza--Klein modes living in the 11th
dimension become tachyonic when this critical line is reached.

In conclusion, the entropy and arrow-of-time problems are neatly solved, in PBB cosmology,
by the identification of our observable Universe with the interior of a large, primordial
 black hole. The entropy of the black hole is large, because of its  size ($>10^{20}l_s$)  
and, therefore, as with other features of PBB cosmology, this can be objected to 
 as  huge fine-tuning \cite{TW}. My answer to this objection, as to
the others, is simple: the classical collapse/inflation process is scale-free; it  should
therefore lead to a flattish distribution of horizon sizes, extending from a minimal
stringy size to very large ``macroscopic" scales.
Given such a size, no other ratio is tuned to a particularly large or small
value. Next, there is a built-in mechanism to provide
saturation of the HEB till the end of the DDI phase, and for the HEB to
force an exit to the radiation-dominated FRW phase.
From there on, the entropy budget story is simple: our entropy remains, to date,
 roughly constant and around
$10^{90}$, while $S_{BH}$ keeps increasing --at somewhat different rates-- during the radiation-
 and  matter-dominated epochs,  reaching about $10^{120}$ today.
Thus our entropy has still a long way to go while it keeps fixing our arrow-of-time!

\vspace{5mm}

It is a pleasure to thank the organizers of this meeting for their kind invitation
and to wish Fran\c cois many more years of highly rewarding research.

\begin{figure}
\hglue 1cm
\epsfig{figure=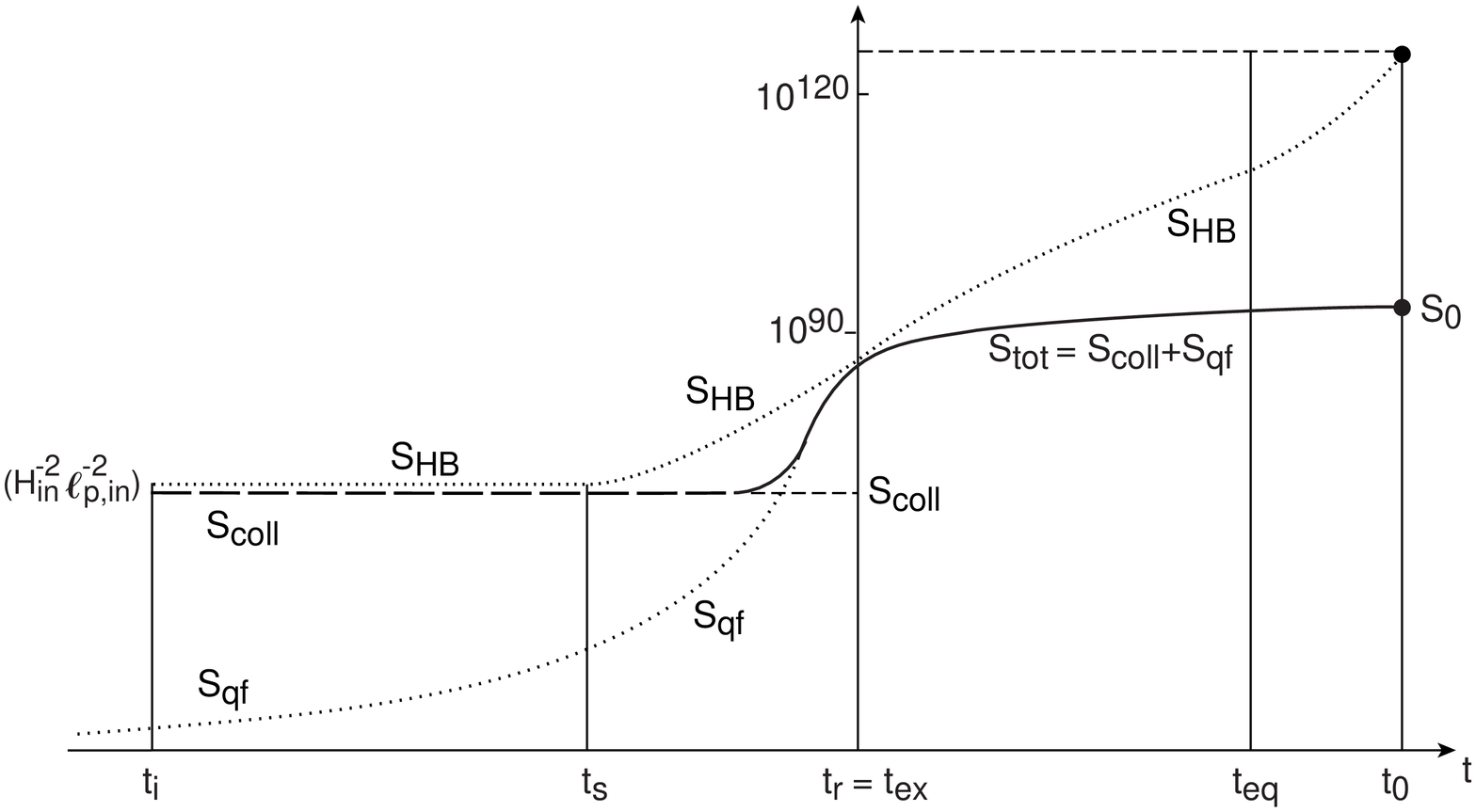,width=12cm}
\caption[]{Entropy history from the onset of PBB inflation till today.
Various contributions to the entropy budget of our Universe are shown together with
the Hubble entropy bound $S_{HB}$. See  text for explanation.}
\end{figure}

\begin{figure}
\hglue 1cm
\epsfig{figure=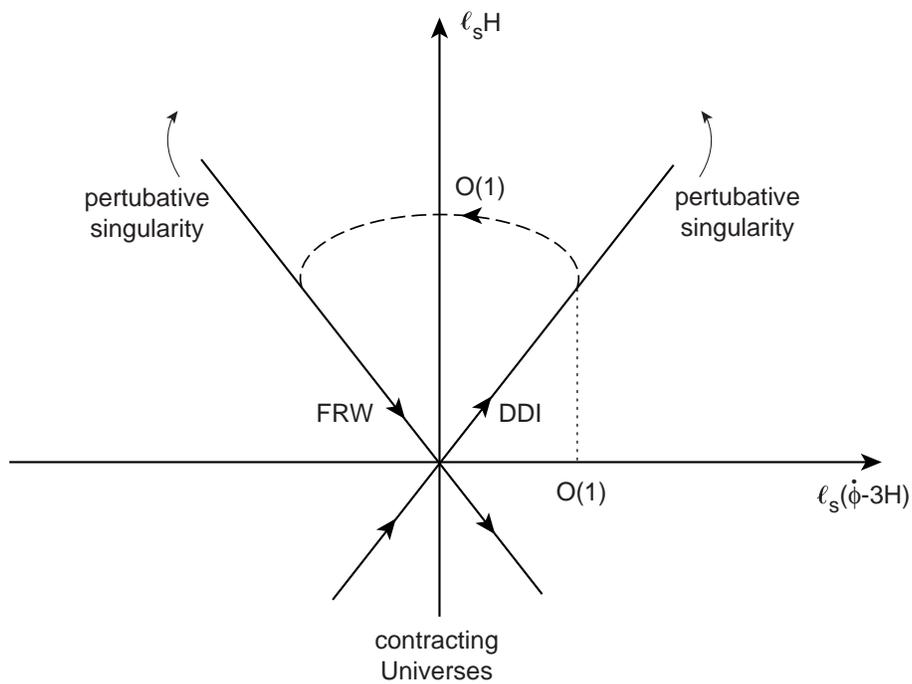,width=12cm}
\caption[]{The standard phase diagram of string cosmology showing the duality-related
inflationary and FRW branches and a possible (dashed) path connecting the two}
\end{figure}

\begin{figure}
\hglue 1cm
\epsfig{figure=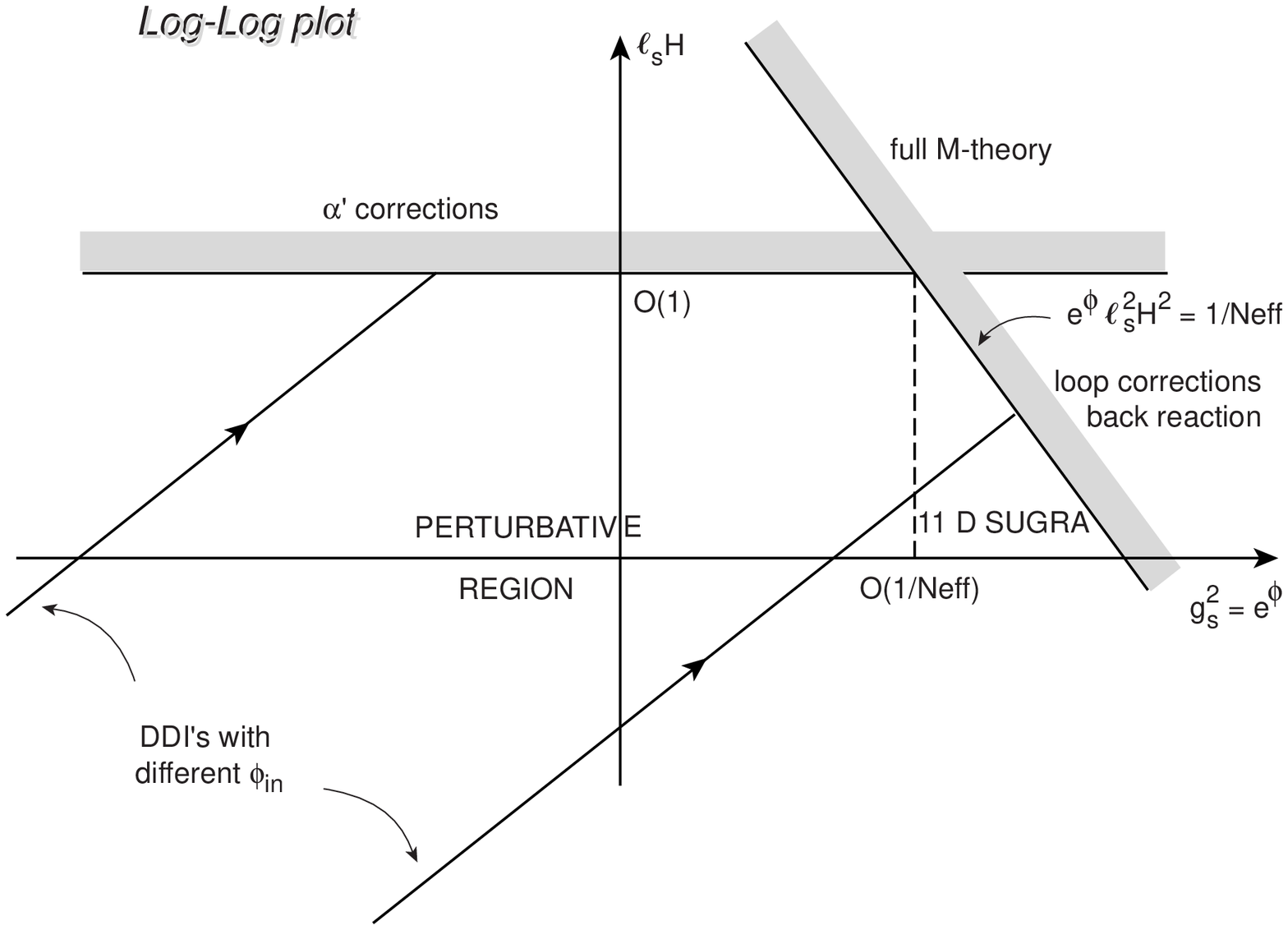,width=12cm}
\caption[]{Another phase diagram showing different perturbative solutions as they hit
boundaries where either curvature or coupling become large, depending on initial data}
\end{figure}


\newpage
 


\begin{thebibliography}{99}

\newcommand{\plb}{{\em Phys. Lett. B}\ }
\newcommand{\prl}{{\em Phys. Rev. Lett.}\ }
\newcommand{\prd}{{\em Phys. Rev. D}\ }
\newcommand{\npb}{{\em Nucl. Phys. B}\ }
\newcommand{\bb}{\bibitem}

\bibitem{Penrose}  R. Penrose, {\em The Emperor's new mind} 
(Oxford University Press, New York, 1989), Chapter 7.

\bb{BB} J. D. Bekenstein, {\em Phys. Rev.} {\bf D23} (1981) 287; {\bf D49} (1994) 1912, 
and references therein.

\bb{holography} G. 't Hooft, {\em Abdus Salam Festschrift: a collection of talks}, eds.
 A. Ali, J. Ellis and S. Randjbar-Daemi  (World Scientific, Singapore, 1993), gr-qc/9321026; \\
L. Susskind, {\em J. Math. Phys.} {\bf 36} (1995) 6377, and references therein.

\bb{GVDivonne}  G. Veneziano, in 
{\em Hot Hadronic Matter: Theory and Experiments}, Divonne, June 1994,
eds. J. Letessier, H. Gutbrod and J. Rafelsky,  NATO-ASI Series B: Physics, 
{\bf 346} (Plenum Press, New York 1995),
 p. 63.


\bb{GVFC} G. Veneziano, {\em Europhys. Lett.} {\bf 2} (1986) 133.

\bb{Bekenstein2} J. D. Bekenstein, {\em Int. J.  Theor. Phys.} {\bf 28}  (1989) 967.


\bb{FS} W. Fischler and L. Susskind, {\em Holography and cosmology}, hep-th/9806039.

\bb{Rama} S. K.  Rama and T. Sarkar, {\em Phys. Lett.} {\bf B450 } (1999) 55,
 hep-th/9812043. 

\bb{KL} N. Kaloper and A. Linde, {\em Cosmology vs. Holography}, hep-th/9904120.

\bibitem{inflation} E. W. Kolb and M. S. Turner, {\em The early Universe} 
(Addison-Wesley, Redwood City, CA, 1990); 
A.D. Linde, {\em Particle physics and inflationary cosmology} 
(Harwood, New York, 1990).

\bb{Rey1}  D. Bak and S.-J. Rey, {\em Holographic principle and string cosmology},
hep-th/9811008.

\bb{Maharana} A. K. Biswas, J. Maharana and R.K. Pradhan, {\em The holography
principle and pre-big bang cosmology}, hep-th/9811051.

\bb{Bousso} R. Bousso, {\em A Covariant Entropy Conjecture}, hep-th/9905177;
{\em Holography in General Space-Times}, hep-th/9906022.

\bb{Carr}  B. J. Carr and S. W. Hawking, {\em Mon. Not. Roy. Astr. Soc.}
{\bf 168} (1974) 399; B. J. Carr, {\em Astrophys. J.} {\bf 201} (1975) 1; 
I. D. Novikov and A. G. Polnarev, {\em Astron. Zh.} {\bf 57} (1980) 250 [{\em
Sov. Astron.} {\bf 24} (1980) 147].



\bb{GVHB} G. Veneziano, {\em Pre-bangian origin of our entropy and time arrow},  hep-th/9902126.

\bb{others} R. Easther and D. A. Lowe, 
{\em Holography, Cosmology, and the Second Law of Thermodynamics}, hep-th/9902088; 
D. Bak and S.-J. Rey, {\em Cosmic Holography}, hep-th/9902173.

\bb{Ramy} R. Brustein, {\em The Generalized Second Law of Thermodynamics in Cosmology}, 
gr-qc/9904061.

\bb{GV} G. Veneziano,  {\em Phys. Lett.} {\bf B406 } (1997) 297;
 A. Buonanno, K.A. Meissner,  C. Ungarelli 
and G. Veneziano, {\em Phys. Rev.} {\bf D57} (1998) 2543.

\bibitem{BDV} A. Buonanno, T. Damour and G. Veneziano, {\em Nucl. Phys.} {\bf B543}
(1999) 275, hep-th/9806230. 

\bb{Erice} G. Veneziano, {\em Inflating, warming up, and probing
the pre-bangian Universe},  hep-th/9902097.

\bb{HP} S. W. Hawking and R. Penrose, {\em Proc. Roy. Soc. Lond.} {\bf A314} (1970) 529.

\bibitem{PBB} G. Veneziano,  {\em Phys. Lett. }  {\bf B265} (1991)  287;
 M. Gasperini and G. Veneziano,
{\em Astropart. Phys.} {\bf 1} 
(1993) 317.  An updated collection of papers on the pre-big bang
scenario is available at {\tt Ðhttp://www.to.infn.it/\~{}gasperin/}.

\bibitem{exit} R. Brustein and G. Veneziano, {\em Phys. Lett.} {\bf B329} (1994) 429; 
N. Kaloper, R. Madden and K.A. Olive, {\em Nucl. Phys.} {\bf B452} (1995) 677, 
{\em Phys. Lett.} {\bf B371} (1996) 34; R. Easther, K. Maeda and D. Wands, {\em Phys. Rev.}
 {\bf D53} (1996) 4247;
M. Gasperini, M. Maggiore and G. Veneziano,  
 {\em Nucl. Phys.} {\bf B494} (1997) 315;
R. Brustein and R. Madden, {\em Phys. Lett.} {\bf B410} (1997) 110,
 {\em Phys. Rev.} {\bf D57} (1998) 712.
 
\bibitem{MR} M. Maggiore and A. Riotto, {\em D-branes and cosmology}, hep-th/9811089.
 

\bb{BBB} G. Veneziano, in {\em Effective theories and fundamental interactions},
Erice, 1996, ed. A. Zichichi (World Scientific, Singapore, 1997), p. 300; 
A. Buonanno, K. A. Meissner, C. Ungarelli and G. Veneziano, {\em  
JHEP01}, 004 (1998).

\bibitem{entropy} M. Gasperini and M. Giovannini, {\em Phys. Lett.} {\bf B301} (1993) 334,
{\em Class. Quant. Grav.} {\bf 10} (1993) L133;
 R. Brandenberger, V. Mukhanov and T. Prokopec,  {\em Phys.
Rev. Lett.} {\bf 69} (1992)  3606, {\em Phys. Rev.}  {\bf D48} (1993) 2443.


\bibitem{TW} M. Turner and E. Weinberg, {\em Phys. Rev.} {\bf D56} (1997) 4604; 
  N. Kaloper, A. Linde and  R. Bousso, {\em Phys. Rev.} {\bf D59} (1999) 043508.

\end{thebibliography}
\end{document}